\documentclass[12pt,preprint]{aastex}

\newcommand{\hk}{\mathrm{H}-\mathrm{K}}
\newcommand{\jh}{\mathrm{J}-\mathrm{H}}
\newcommand{\ejh}{\mathrm{E}(\mathrm{J}-\mathrm{H})}
\newcommand{\ehk}{\mathrm{E}(\mathrm{H}-\mathrm{K})}
\newcommand{\hks}{\mathrm{H}-\mathrm{K}_s}
\newcommand{\ehks}{\mathrm{E}(\mathrm{H}-\mathrm{K}_s)}

\shorttitle{Structure of L694-2}
\shortauthors{Harvey et al.}
\begin{document}

\title{Envelope Structure of Starless Core L694-2 Derived from a 
Near-Infrared Extinction Map}

\author{Daniel W.A.\ Harvey,
        David J.\ Wilner,
        Charles J.\ Lada,
        Philip C.\ Myers}

\email{dharvey, dwilner, Clada, pmyers@cfa.harvard.edu}

\affil{Harvard-Smithsonian Center for Astrophysics, 60 Garden Street,
Cambridge, MA 02138}

\and

\author{Jo\~{a}o F.\ Alves}

\email{jalves@eso.org}

\affil{European Southern Observatory, Karl-Schwarzschild Stra{\ss}e 2,
D-85748 Garching bei M\"{u}nchen, Germany}

\begin{abstract}
We present a near-infrared extinction study of the dark globule L694-2,
a starless core that shows strong evidence for inward motions in 
the profiles of molecular spectral lines. The J,H, and K band data 
were taken using the European Southern Observatory New Technology Telescope. 
The best fit simple spherical power law model has index $p=2.6 \pm 0.2$, 
over the $\sim 0.036$--0.1~pc range in radius sampled in extinction.
This power law slope is steeper than the value of $p=2$ for a singular 
isothermal sphere, the initial condition of the inside-out model for
protostellar collapse. Including an additional component of extinction 
along the line of sight further steepens the inferred profile. 
A fit for a Bonnor-Ebert sphere model results in a super-critical value 
of the dimensionless radius $\xi_{\mathrm{max}}=25 \pm 3$. This unstable
configuration of material in the L694--2 core may be related to the
observed inward motions. The Bonnor-Ebert model matches the {\em shape}
of the observed density profile, but significantly underestimates the 
{\em amount} of extinction observed in the L694--2 core (by a factor of 
$\sim 4$). This discrepancy in normalization has also been found for the 
nearby protostellar core B335 (Harvey et al.\ 2001). A cylindrical model 
with scale height $H=0.0164 \pm 0.002$~pc ($13.5'' \pm 5''$) viewed at a 
small inclination
to the axis of the cylinder provides an equally good radial profile as a
power law model, and it also reproduces the asymmetry of the L694--2 core
remarkably well. In addition, this model provides a possible basis for 
understanding the discrepancy in the normalization of the Bonnor-Ebert model,
namely that L694--2 has prolate structure, with the full extent (mass) of 
the core being missed by analysis that assumes symmetry between the profiles
of the core in the plane of the sky and along the line-of-sight. If the core
is sufficiently magnetized then fragmentation may be avoided, and later 
evolution might produce a protostar similar to B335.
\end{abstract}


\section{Introduction}

Molecular line surveys of nearby dark clouds have identified a large sample 
of dense cores (e.g.\ Benson \& Myers 1989), of which roughly half are 
associated with young stellar objects detected by the IRAS (Beichman et al.\ 
1986). The properties of these cores have been adopted in 
the standard model of isolated star formation whereby a slowly rotating, 
nearly spherical core makes a star/disk system via ``inside-out'' collapse
(Shu, Adams \& Lizano 1987). In this picture, a ``starless'' dense core
represents the earliest stage of the star formation process. The physical
conditions in this early stage have a profound impact on the evolution of
protostars towards the main sequence. The initial density structure,
particularly in the innermost regions, affects the collapse dynamics and
the time dependence of the mass accretion rate (Foster \& Chevalier 1993),
and therefore many of the observable properties of protostars, including
luminosity.

While the standard theory has proved very successful in many predictions    
(see the review of Shu et al.\ 1993), its weakest foundation arguably 
lies in the adopted initial conditions (Andr\'{e}, Ward-Thompson \& Barsony
2000). In the simplest case where a spherical starless core loses turbulent
and magnetic support and relaxes to a balance between gravity and 
thermal pressure, an $r^{-2}$ density distribution is established and the
core collapses from the inside-out with a constant mass accretion rate (Shu 
1977). Alternatively, if collapse begins before the density distribution  
fully relaxes, then a central region of relatively constant density remains 
and the mass accretion rate is an order of magnitude larger at early times 
(Foster \& Chevalier 1993). This phenomenon has been identified with the   
youngest ``Class~0'' protostars, which exhibit especially powerful outflows 
(Henriksen, Andre \& Bontemps 1997; Andre, Ward-Thompson \& Barsony 2000). 
In another case, a starless core with a logotropic equation of state  
develops a nearly $r^{-1}$ density law and the mass accretion rate during 
collapse increases rapidly with time (McLaughlin \& Pudritz 1997).

Better observations of starless cores are needed to measure appropriate
initial conditions. Molecular hydrogen, the main mass constituent, 
cannot be observed directly in these cold (T$<$10~K) dense regions.
The most robust tracer of mass is provided by dust. (Molecular line 
emission is compromised by excitation effects and severe depletion.) 

Most of the detailed information on starless core structure comes 
from observations of dust emission, using data from submillimeter 
bolometer cameras (e.g. Ward-Thompson, Motte \& Andre 1999,
Shirley et al. 2000, Visser, Richer \& Chandler 2001).
The intensity of the dust emission provides an integral along the 
line-of-sight of the product of the dust temperature and density. 
An important conclusion from these studies is that starless 
cores appear to show flat density profiles in their inner regions, 
with extended envelopes that fall off rapidly in power law fashion.
However, the regions of relatively constant density are poorly 
resolved by the low angular resolution available in the submillimeter.
Moreover, the interpretation of the flat central density gradients 
has been called into question: more sophisticated analysis including 
self-consistent temperature calculations predict much smaller regions of 
flattening, or no flattening at all, in large part because the cores are   
cooler in their deep interiors than assumed previously (Evans et al.\ 2001,
Walmsley et al.\ 2001).

Observations of dust extinction provide a more reliable measure of 
column density than observations of dust emission, mainly because 
extinction is not sensitive to gradients of dust temperature. 
Historically, this method has been impossible to apply to small regions 
of high extinction because of poor sensitivity and a paucity of background 
stars, even in the infrared where the extinction is much less than 
in the optical (e.g.\ Bok \& Cordwell 1973, Jones et al.\ 1980, 1984). 
The development of large format near-infrared array cameras has sparked 
renewed interest in the technique (Lada et al.\ 1994, Alves et al.\ 1998,
Alves, Lada \& Lada 1999, 2001, Harvey et al. 2001).

Harvey et al.\ (2001) describe deep near-infrared extinction measurements 
toward B355, arguably the best protostellar collapse candidate, using the
Hubble Space Telescope (HST). The B335 system contains a deeply embedded 
young stellar object, and exhibits molecular line profiles well fit by 
inside-out collapse (Zhou et al.\ 1993, Choi et al.\ 1995). The shape of the 
density profile   
determined from the deep extinction data provide the best evidence 
yet in support of inside-out collapse theory. Alves, Lada \& Lada (2001) 
used the color excess method to study the starless core B68 with data from 
the the European Southern Observatory (ESO) New Technology Telescope (NTT). 
The B68 system is much less centrally condensed than B335, with 
lower peak column density, and it was easily penetrated with the 
sensitivity available from the ground. The extinction map of B68 
suggests a hydrostatic equilibrium structure, with density distribution 
best described by the equations of a pressure confined, self-gravitating, 
isothermal sphere (Ebert 1955, Bonnor 1956).

The examples of B68 and B335 represent milestones in the early evolution 
of low-mass stars, but they do not show the density structure of a 
forming dense core, or the density structure at the onset of collapse. 
The recent identification of contracting starless cores presents 
an opportunity to study this important intermediate evolutionary phase. 
There are now at least 7 ``strong'' infall candidates among the 
starless cores, based on observations of molecular line profiles with 
redshifted self-absorption in a systematic search of more than 200 targets 
(Lee, Myers \& Tafalla 1999). Molecular line maps of these objects provide
strong evidence of inward motions, with speed $\sim 0.1$~km~s$^{-1}$ over
a radius of $\sim 0.1$~pc (Lee, Myers \& Tafalla 2001). The physical basis
for these motions is unclear. The speeds are subsonic and may be associated
with condensation through ambipolar diffusion, or perhaps a magnetically
diluted gravitational collapse (Ciolek \& Basu 2000).
Alternatively, pressure driven motions due to the dissipation of turbulence
may be responsible (Myers \& Lazarian 1998). 

L694--2 is the best of these sources for extinction work, since it is viewed 
against a very high density of background stars near the Galactic plane 
($l=44.9^{\circ}$, $b=-6.6^{\circ}$), and at the same time there is
no significant reddening by additional clouds along this line-of-sight.
This isolated, round, dense core is situated close to the protostar B335   
in the sky, probably associated with the same molecular cloud complex. This
association suggests a distance of 250~pc (Tomita, Saito \& Ohtani 1979),
consistent with recent estimates of $230 \pm 30$~pc (Kawamura et al.\ 2001).
But unlike B335, L694--2 is starless, with no evidence of an embedded 
luminosity source from IRAS or any indication of bipolar outflow (Visser 
2000, Harvey et al.\ 2002). Analysis of dust emission suggests a steep outer
profile with flattening of the density gradient within a radius of a few 
thousand AU (Visser 2000), but suffers from the probably inaccurate 
assumption of an isothermal temperature distribution (Evans et al.\ 2001)
In this paper we present a near-infrared color excess study of L694--2, using
observations made at the ESO NTT. These data probe the density 
distribution between radii of $0.03$~pc (6000~AU) to the outer edge 
of the globule at $\sim 0.15$~pc, where the core merges into a more extended 
distribution of gas.

\section{Observations and Data Reduction}

\subsection{The Near-Infrared Color Excess Technique}
We provide here a brief review of the Near-Infrared Color Excess technique
for measuring the structure of a dense core. For a more detailed description
of the method, the reader is referred to Lada et al.\ (1994) and Harvey et 
al.\ (2001).

The basic method is to measure the near-infrared color excess for stars 
background to a dense core:
\begin{equation}
\ehk = (\hk)_{\mathrm{observed}} - (\hk)_{\ast} \ ,
\end{equation}
where $(\hk)_{\ast}$ is the intrinsic color of the star. The color excess 
is directly proportional to the dust column density. Thus 
color excess may be converted to gas column density via the near-infrared 
extinction law and an assumed gas-to-dust ratio. These column density 
estimates can be used to determine the overall density structure of the
obscuring core.

In practice, the method does not require a knowledge of
the intrinsic colors of the individual extincted stars. The proximity
of the near-infrared filters to the Rayleigh-Jeans region of the spectrum
for main sequence and giant stars means that the intrinsic colors are small,
$0 < (\hk)_{\ast} < 0.3$ (Koornneef 1983). A statistical
correction to their observed colors may be obtained empirically from
the background stellar population (provided it is sufficiently
spatially uniform).

\subsection{NTT Observations}
Observations of L694--2 were made on 2001 June 8 using the Son of Isaac 
(SofI) infrared camera on the ESO NTT. The SofI camera has 
$1024 \times 1024$ pixels with size $0\farcs29$, giving a $5'$ field of view.
Each basic observation consisted of a mosaic of ten dithered sub-images, 
each with a total exposure time of 30 seconds, made up of 5 co-adds of 6 
seconds. Dithering offsets were randomly chosen within a $40''$ box.
Observations were made with filters J, H, and K$_s$\footnote{
The K$_s$ filter is medium-band version of the standard K filter, that 
avoids both the atmospheric absorption feature at 1.9~$\mu$m and radiation 
from the thermal background beyond 2.3~$\mu$m}. Figure~1 shows a Digital
Sky Survey (DSS) image of the L694--2 region with the NTT observed field 
indicated.

The data reduction and calibration was done in IRAF following the
standard dark/flat/sky subtraction procedures, using a running flat.
The typical seeing during the night was $0\farcs7$.

\subsubsection{Photometry}
Stars were identified in each image using the {\em SExtractor} program.
Photometry was performed for all non-saturated stars using the {\em
apphot} package in IRAF. A series of 3, 4, and 5 pixel apertures was used 
for every star. Aperture corrections were calculated for each aperture based
on the seeing in the image using a regression to aperture corrections 
calculated for 23 previous observations with the NTT (Huard, private 
communication). While the
4-pixel aperture was used to produce the final photometry catalogues, the 
combination of the three measurements was used to eliminate spurious 
detections and blended stars (roughly 4\% at H, K$_s$), based on whether the 
measurements were consistent with the predicted uncertainties at a 
1.5~$\sigma$ level. 

Due to a not perfect alignment of the Large Field objective on the NTT, 
stellar images are radially elongated in the Northern part of the field of 
view. This elongation affects a strip of about 150 pixels and smoothly 
disappears moving toward the center of the field. Although this distortion 
does not create astrometric problems, it does mean that aperture photometry 
with a constant aperture correction will lead to systematic errors in the 
magnitudes of the stars in the affected region. We have dealt with this 
effect by applying additional spatially dependent aperture corrections.
Each image was divided into a $4 \times 4$ grid, and we calculated the ratio
of the median FWHM in each region to the median FWHM in the Southern 
(bottom) half of the image. The results were consistent between the three 
filters, and the most conservative (smallest) value was taken for each
region in the grid. The resulting grid of spatially dependent seeing FWHMs 
was used to calculate the appropriate additional aperture corrections in 
each region. The additional correction are zero in the Southern half of the 
images, and are largest in the four Northern-most regions, where the FWHM of 
the point-spread function is typically 20\% larger. In all cases, the 
corrections are less than 0.06 magnitudes in size for the 4-pixel aperture, 
and the colors are even more robust, with largest correction of only 
$\hks=-0.013$.

The World Coordinate System in the images was fixed using the {\em imwcs} 
routine written by Doug Mink and stars from the US Naval Observatory 
A2.0 Catalog (Monet et al.\ 1998). The coordinate system is accurate to 
within an arcsec in RA and Dec. Finally, correct zeropoints were calculated 
using 182 matches in the 2 Micron All Sky Survey (2MASS) point source 
catalog. The registration with 
the 2MASS zeropoints is accurate to $\sim 1$\% in each filter. The 2MASS 
zeropoints are themselves accurate to a similar level.

The limiting magnitudes are 19.5 at J, 18.9 at H, and 18.5 at K$_s$, 
corresponding to a 0.25 magnitude error limit (signal to noise of 4) and a 
completeness of roughly 80\%. The final photometry catalog contains 
1451 stars detected at both H and K$_s$.

\subsection{Reddening Law and Conversion between CIT and SofI/NTT Colors}
Carpenter (2001) provides extensive 
conversion relations between 2MASS colors and magnitudes with those for
other filter systems. Of particular interest to us here are the conversion
to the ESO filter system (as used for our observations) and the California 
Institute of Technology (CIT) filter
system (as used during our extinction study of B335). Conversion relations
are also presented in Bessell \& Brett (1988). Unfortunately, both the 
Carpenter and Bessell \& Brett conversions are derived over a very small 
range in $\hk$ colors, generally $\hk \lesssim 0.5$, and perhaps can not be 
trusted for the reddest stars in our catalog.

For the purposes of the present study, we divide our attention to 
two issues concerning the photometric systems: 1) how the intrinsic $\hks$ 
colors of stars in our catalog relates to their $\hk$ on the CIT system;
2) how the color-excess $\ehks$ is related to $\ehk$ on the CIT system. The
2MASS conversion relations are sufficient to address (1), while (2) can be
investigated by studying a color-color diagram of the stars in our study.
At the low colors ($\hk < 0.3$) intrinsic to main sequence and giant stars,
the conversions from our instrumental colors to the 2MASS system are
negligibly small compared to the systematic uncertainties ($\lesssim 0.01$),
due to our registration with 2MASS zeropoints. For the intrinsic colors
(ie $\jh < 0.6$, $\hk < 0.3$), 
the transformations between our colors and the CIT system are therefore:
\begin{eqnarray}
(\jh) & = & 1.076 (\jh)_{CIT} -0.043\\
(\hks) & = & 1.026 (\hk)_{CIT} + 0.028
\end{eqnarray}
with an uncertainty of $\sim 0.01$.

Figure~2 shows a color-color diagram for the subset of stars in the catalog 
with signal-to-noise of $>10$ in each filter. The solid lines on the plot 
show the loci of unreddened main sequence and giant stars, while the two 
dashed lines show the expected reddening zone for these stars, using the 
standard reddening relations (Rieke \& Lebofsky 1985), assuming 
$\ejh=1.076\ejh_{CIT}$ 
and $\ehks=1.026\ehk_{CIT}$ --- an extrapolation of the low-color 
transformations. The reddening vectors follow the stellar colors remarkably
well, even out to a color of $\hk \sim 1.5$. This suggests that the
reddening law can be modified easily to a form $A_V/\ehks$ using 
$A_V/\ehks \simeq 1.026*(A_V/\ehk)=16.3$, while introducing little extra
systematic error compared to that intrinsic to the reddening law itself.

\section{Results and Analysis}

\subsection{The Background Population Mean Color and Dispersion}
\label{sec:background}
The DSS image of L694--2 in Figure~1 shows that the 
core is associated with an extended filamentary structure of extinction. 
This lower level extinction extends away from the core towards the 
South-East and the North. To characterize the background population, we 
therefore study the stellar properties in the Western half of the images, 
considering only stars that are more than $2\farcm5$ from the peak of the 
mm-wave emission. This minimum radius was chosen on the basis of minimizing 
any remaining extinction, but retaining enough stars to draw reliable 
conclusions about the background properties. This ``background'' region 
contains a total of 221 stars detected at both H and K$_s$. These stars 
have mean color of $\overline{\hks}=0.35$ and standard deviation of 
$\sigma(\hks)=0.21$.

The mean color is somewhat larger than expected for an unreddened stellar
population, since the intrinsic colors of main sequence and giant stars span
the narrow range $0< \hk <0.3$, and suggests that the stars in question are
not entirely free from extinction.
This conclusion is also supported by comparison with our results for the
stellar background near B335, a globule that lies very close to L694 on the
sky ($< 4^{\circ}$ away), and is probably even associated with the same 
parent molecular cloud. The mean color and standard deviation for the 
stellar population at a distance of between $2\farcm5$ and 3' from the B335 
center are $\overline{\hk}=0.26$ and $\sigma(\hk)=0.21$, which transform to
$\overline{\hks}=0.29$ and $\sigma( \hks)=0.22$. Those values are very 
similar to the results found here. Off field images in the B335 vicinity 
showed that the true background properties became $\overline{\hk}=0.13$ and 
$\sigma(\hk)=0.16$, corresponding to $\overline{\hks}=0.16$ and 
$\sigma(\hks)=0.17$. As a test for similarity between the L694 and B335
background, we have compared the luminosity functions for the L694 
``background'' with the deep H-band luminosity function that was calculated 
in the B335 study ($N(m<H)=5.4 (H/14.17)^{9.56}$ stars arcmin$^{-2}$, for 
$H>14.17$). The luminosity functions are entirely indistinguishable to
within the counting noise. We therefore conclude that the B335 background 
properties represent a reliable measure of the background to L694, and adopt
those properties in favor of the values calculated for the stars near the
Western edge of the L694 image.

\subsection{Color Excess Distribution near L694--2}
Figure~3 conveys much of the information from the NTT/SofI observations of
L694--2. The left-hand plot is a pseudoimage in that the two axes are 
spatial, the apparent brightness of a star at H-band determines its ``size'',
and the value of its $\hks$ color determines its ``color''. The upper 
right-hand plot displays the radial dependence of the $\hks$ colors out to 
$200''$ (0.24~pc at 250~pc distance) from the center. The lower right-hand 
plot shows the 
radial colors convolved with an annular Kernel that is a Gaussian of 
logarithmic width 10\% in radius, in order to better highlight the profile 
of the $\hk$ colors. Superposed are the standard deviation (dashed line), 
and standard error (dotted line) at each convolved point, the former 
indicating the spread of colors at each radius, and the latter indicating 
the uncertainty in the mean color at each radius.
The origin in every plot is given by the position of peak emission
in the N$_2$H$^+$(1--0) spectral line measured at the Berkeley Illinois 
Maryland Array (BIMA) by Jonathon Williams and collaborators 
(unpublished observations): R.A.=19:41:04.44, Dec.=10:57:00.9 (J2000). The
uncertainty in the position of this peak is around $2''$. As can be seen from
the pseudoimage, the region of highest column density is extended towards 
the South-East corner of the image. The region is marked
in the image with dashed lines. This structure is also visible in the 
DSS image of the region in Figure~1, and is associated with an extended
filamentary structure of gas. To prevent confusion, stars within this 
region have been omitted from the radial-color plot. Of the 1451
stars in the image detected at H \& K with $S/N > 4$, 199 stars have impact
parameters smaller than $83''$ or $0.1$~pc, of which 14 lie in the ``bar'' 
region. The detected star closest to the N$_2$H$^+$ peak is at $28''$ 
(0.034~pc) distance. This star is also the reddest in the sample and has 
color $\hks=2.17$, corresponding to over 30~mag of equivalent visual 
extinction. In addition there is a star at impact parameter $30''$ (0.036~pc)
that is well detected at K$_s=16.05\pm0.03$, but too red to be detected at H.
Based on the assumed H-band limit of 18.9, where the magnitude error is 
$\sim 0.25$, this star provides a lower limit of $\hks > 2.81 \pm 0.25$ 
($A_V >40$).

The major features of Figure~3 are: (1) a strong gradient in the $\hk$ colors
as one approaches the N$_2$H$^+$ peak; (2) many fewer stars detected close 
to the center; (3) a central region of column density that is so high that
it cannot be penetrated; (4) reddening in the stars changing behavior and 
flattening off beyond about $83''$ or $0.1$~pc, apparently the boundary/edge 
of the dense core and a more extended distribution of gas; and (5) regions of
high color excess that extend to the image edge in the South-East, and 
part-way to the image edge in the North --- the shape of these regions match 
the profile of L694--2 in the DSS image, and highlight the 
somewhat filamentary nature of the L694 region.

\subsection{Theoretical Models of the L694--2 Density Distribution}
We describe several theoretical models for the density distribution of
L694--2 and evaluate the success of these models in light of the
extinction data. The models considered are not meant to comprise an
exhaustive list. They include Bonnor-Ebert spheres, inside-out collapse, 
simple power law descriptions, and Plummer-like models. We also discuss 
filamentary models in the context of departures from spherical symmetry.

\subsubsection{Inside-Out Collapse Models}
There is a substantial literature of theoretical work by Frank Shu and
colleagues describing the density structure of collapsing dense cloud
cores, starting from the initial condition of a singular isothermal
sphere (SIS). The density distribution of the SIS falls off as
$\rho \propto r^{-2}$, where $r$ is the radius, with the
normalization of the density determined by the effective sound speed.
For L694--2, Lee et al.\ (2001) have measured the FWHM of the N$_2$H$^+$
line to be $0.25$~km~s$^{-1}$. If we assume a central temperature of 9~K
based on the models of Evans et al.\ (2001), then this suggests an effective
sound speed of  $a=0.20$~km~s$^{-1}$ with a turbulent component of 
$a_{\mathrm{turb}}=0.093$~km~s$^{-1}$. The static initial condition in
the inside-out collapse model is therefore given by: 
\begin{eqnarray}
\rho_{\mathrm{static}}(r) & = & \frac{a^2}{2 \pi G r^2}  \\
& = & 1.0 \times 10^{-20} \; \mathrm{g} \, \mathrm{cm}^{-3} \; 
      \left(\frac{a}{0.20 \, \mathrm{km} \, \mathrm{s}^{-1}} \right)^2 
      \left(\frac{r}{0.1 \, \mathrm{pc}} \right)^{-2} \\ 
n_{H_2}(r) & = & 2.2 \times 10^{3} \; \mathrm{cm}^{-3} 
\; \left(\frac{a}{0.20 \, \mathrm{km} \, \mathrm{s}^{-1}} \right)^2 
\left(\frac{r}{0.1 \, \mathrm{pc}} \right)^{-2} \ , 
\end{eqnarray}
where the conversion to a molecular hydrogen number density assumes
a mean molecular weight of 2.3.

The simplest scenario is the spherically symmetric collapse (Shu
1977). In this model, a spherical wave of collapse propagates outwards
from the center at the effective sound speed. The radial distance that
the wave has traveled, sometimes called the infall radius, is the only
additional parameter that defines the density distribution. Inside the
infall radius, conditions approach free fall, with the density taking
the asymptotic form $\rho \propto r^{-3/2}$. Harvey et al.\ (2001) 
describe deep near-infrared extinction measurements toward B355, arguably 
the best protostellar collapse candidate. This system, which contains a 
deeply embedded young stellar object, exhibits molecular line profiles 
well fit by inside-out collapse (Zhou et al.\ 1993, Choi et al.\ 1995). 
The shape of the density profile determined from the deep extinction 
data provide the best evidence yet in support of inside-out collapse 
theory, in particular the $r^{-2}$ fall off for the envelope and inner 
turnover towards free-fall. Recent observations of dust emission from B335 
using mm-wave interferometry have shown that the inner density
distribution is very close to $\rho \propto r^{-3/2}$, although the 
way in which the density asymptotes to this behavior does not quite agree
with the exact details of the inside-out collapse model (Harvey et al.\ 
2003).
 
In the inside-out collapse model, a central point source with luminosity 
$L \sim a^6 t / G R_{\ast}$ is formed. For L694--2, the limits on the
luminosity of any embedded point source of $L \lesssim 0.3$~L$_{\odot}$,
combined with the extended nature of the inward motions inferred from 
molecular line mapping (Lee et al.\ 2001), are inconsistent with the
inside-out collapse model. Nevertheless we consider this model due to 
its success in describing the density structure of B335.

\subsubsection{Bonnor-Ebert Models}
Bonnor-Ebert models are pressure-confined isothermal spheres, for
which the solution remains finite at the origin (Ebert 1955, Bonnor
1956). In common with the singular isothermal sphere, the initial
condition for inside-out collapse, they are solutions of a modified
Lane-Emden equation (Chandrasekhar 1967):
\begin{equation}
\frac{1}{\xi^2} \frac{d}{d \xi} \left( \xi^2 \frac{d \psi}{d \xi}
\right) = \exp{(-\psi)} \ ,
\end{equation}
where $\xi=(r/R_0)$ is the dimensionless-radius,  
$R_0=a/\sqrt{4 \pi G \rho_c}$ is the (physical) scale-radius, and 
$\psi( \xi)=- \ln{ (\rho/ \rho_c )}$ is a logarithmic
density contrast, with $\rho_c$ the (finite) central density. Unlike
the singular solution, the Bonnor-Ebert solutions do not diverge at
the origin. Instead, the boundary conditions are that the function
$\psi$ and its first derivative are zero at the origin.

The above equation can be solved by division into two first order
equations which can then be tackled simultaneously using numerical
techniques (in our case a 4th order Runge-Kutta method). There is a
family of solutions characterized by a single parameter --- the
dimensionless outer radius of the sphere, $\xi_{\mathrm{max}}$. For a
given sound speed and a particular choice of the shape of the density
curve (i.e.\ $\xi_{\mathrm{max}}$), there is one additional degree of
freedom: the physical scale of the model. This additional degree of
freedom has often been removed by implementing an additional constraint, 
for example by fixing the outer radius, or the total mass of the globule.
However, if such a constraint is applied, it is important to make some 
estimate of its uncertainty, and take that uncertainty into account in the 
fitting process in order to properly interpret both the systematic and 
random errors in $\xi_{\mathrm{max}}$.

Configurations with dimensionless outer radius $\xi_{\mathrm{max}} >
6.5$ are unstable to gravitational collapse (Bonnor 1956). The
gravitational collapse of Bonnor-Ebert spheres has been studied
numerically by Foster \& Chevalier (1993). They find that collapse begins
in the flat inner region, with the peak infall velocity at the 
``shoulder'' of the density distribution. The flow
asymptotically approaches the Larson-Penston solution at the origin at
the time of and prior to the formation of a central core, as material from
the flat region collapses into the center. These large early accretion rates
last for a short time. If the cloud is initially very centrally condensed 
(i.e.\ $\xi_{\mathrm{max}} \gg 6.5$) the later stages of infall closely
resemble Shu's inside-out collapse of a singular isothermal sphere.

Recently, Alves, Lada \& Lada (2001) used the color excess method to
study B68, a starless core, with data from the ESO NTT.  The B68
system is much less centrally condensed than L964-2, and has much lower
column density at the center (only about 30 magnitudes of equivalent
visual extinction). The Alves et al.\ (2001) extinction map of B68
suggests that the density structure is well described by a
Bonnor-Ebert sphere with dimensionless outer radius slightly in excess
of critical: $\xi_{\mathrm{max}}=6.9 \pm 0.2$. 
A highly super-critical Bonnor-Ebert sphere was found to fit 
the near-infrared extinction in protostellar collapse candidate B335 
(Harvey et al.\ 2001). The model, with $\xi_{\mathrm{max}}=12.5 \pm 2.6$ 
was indistinguishable from the inside-out collapse model over the range
in radius where stars were detected, $r\gtrsim 0.017$~pc or $14''$.
Subsequent observations of dust emission with IRAM PdBI have shown power-law
behavior ($p \sim 1.5$) in the inner region and rule out the Bonnor-Ebert 
model. However, the Bonnor-Ebert sphere is a hydrostatic model, and since
the central regions of B335 are clearly not in equilibrium (presence of
YSO, outflow, infall), it is not surprising that the model does not
accurately describe the inner density structure of B335. The physical 
interpretation of the inward motions detected in L694--2 by Lee et al.\ 
(2001) is unclear. If the motions represent a phase of condensation, then
the density structure of the core may be well described by a supercritical
Bonnor-Ebert model.

\subsubsection{Power Law models}
The singular isothermal sphere represents a special case of the family of
power law models for density structure: $\rho(r) \propto r^{-p}$. Other 
special cases include the free-fall model ($p=1.5$) and the logatropic
model ($p=1$), where the gas is governed by a logatropic equation of state
(McLaughlin \& Pudritz 1997).

Visser (2000) fitted a broken power law model to SCUBA observations of 
L694--2. The radial profile of the dust emission from the core was fit
by an isothermal model that had a flat gradient in the inner regions 
($\rho \propto r^{-0.8}$ for $r < 0.04$~pc or $32''$), surrounded by a steep
$\rho \propto r^{-2.7}$ envelope. No uncertainty in these power-law indices
is quoted, but nevertheless the profile represents a very unstable 
arrangement of material, and is consistent with the large inward motions 
detected by Lee et al.\ (2001). The analysis is based on an assumed constant
temperature distribution. However, recent modeling suggests the dust 
temperature will decrease towards the center of a starless core (Evans 
et al.\ 2001). This decrease in the temperature may be what causes 
the flattening in the emission profile in the inner regions, and a density 
profile that does not turn-over at all cannot be ruled out. The turnover
radius measured by Visser should perhaps be construed as an outer limit,
since a more physical temperature profile would inevitably produce a smaller
value.

\subsubsection{Plummer-like Models}
The Plummer-like model is an empirical model suggested by Whitworth \&
Ward-Thompson (2001) that captures the essential observed properties of
pre-stellar cores with a minimum of free parameters. These properties 
include: 1) density distributions that combine flat inner profiles 
($\rho \sim $constant, to $\rho \propto r^{-1}$) with steep outer profiles 
(up to $\rho \propto r^{-4}$); 2) dynamical timescales that reproduce the
short-lived (few$\times 10^4$~yr) high accretion-rate 
($\gtrsim 10^{-5}$~M$_{\odot}$~yr$^{-1}$) nature of the class~0 protostellar
phase, and the longer-lived (few$\times 10^5$~yr) slower accretion-rate
($\lesssim 10^{-6}$~M$_{\odot}$~yr$^{-1}$) nature of the class~I phase; 3)
extended and roughly uniform velocity fields, $v \sim 0.1$~km~s$^{-1}$ over 
$r \sim 0.1$~pc ($83''$).

The model assumes that when a prestellar core becomes unstable against 
collapse at time t=0, it is static and approximates to a Plummer-like density
profile (Plummer 1911), of the form:
\begin{equation}
\rho(r, t=0)=\rho_{\mathrm{flat}} \left[ 
\frac{R_{\mathrm{flat}}}{(R_{\mathrm{flat}}^2+r^2)^{1/2}} \right]^{\eta}
\end{equation}
The initial density is therefore uniform for $r \ll R_{\mathrm{flat}}$, and
falls off as $r^{-\eta}$ for $r \gg R_{\mathrm{flat}}$. This is similar to 
the broken-power law model fitted to L694--2 by Visser (2001).

Subsequent evolution of the model is calculated assuming a negligible 
internal pressure, with the cloud therefore undergoing free-fall collapse.
Collapse proceeds analogously to the Foster \& Chevalier simulations of
collapsing Bonnor-Ebert spheres: a large initial accretion rate as material 
from the uniform density region collapses onto the origin (associated with the 
Class~0 phase), followed by a more sedate accretion of the steep envelope.
During the collapse process the central density profile steepens, 
asymptotically approaching $r^{-3/2}$, but the outermost profile stays 
virtually unchanged. The velocity distribution is at early times peaked at
the shoulder of the density distribution, but the inward motions in the 
central regions eventually reaches and exceeds this value after one free-fall
time.

Whitworth \& Ward-Thompson (2001) propose a fixed value of $\eta=4$ in the
model in order to reproduce the relative lifetimes and accretion rates for 
the Class~0 and Class~I phases, and point out that for $\eta \leq 3$ the 
mass of the protostellar core will diverge due to an infinite reservoir of 
mass. However, in reality any core is likely to be pressure-truncated at 
some point by its interaction with the interstellar medium (c.f.\ 
Bonnor-Ebert sphere), which circumvents this theoretical problem with 
$\eta \leq 3$. 
Moreover, the observed dust-emission profiles of L694--2 (Visser 
2001), and starless core L1544 (Lee et al.\ 2003) are better reproduced with
a value of $\eta=3$.

The model is highly optimistic in that it attempts to reproduce the density
and velocity structure of pre-stellar, Class~0 and Class~I sources. The model
provides a useful framework for interpreting observations of pre-stellar
cores, particularly of the inward velocity fields which cannot be explained 
in the context of a true (hydrostatic) Bonnor-Ebert sphere. In the case of 
Class~0 objects, the model is unable to reproduce the density structure of 
B335, the best studied of such sources, which has outer density profile that 
is very close to that of an isothermal sphere $\rho \propto r^{-2}$, and is 
significantly more shallow than that of the Plummer-like model.

\subsubsection{Filamentary Models}
The complex containing L694--2 provides an example of filamentary structure
that is common to many molecular cloud complexes (see e.g.\ Schneider \&
Elmegreen 1979). Such structure can be produced by a variety of physical
processes. Mechanisms that are well known to trigger star
formation, such as cloud-cloud collisions, compression by a shell of a 
supernova remnant, and encounters with an ionization front surrounding an
OB star all produce sheet-like cloud structures. Magnetic support might also
result in an initially flattened cloud (e.g.\ Mouschovias 1976).
Filamentary structures will probably result from the fragmentation of such 
sheet-like clouds (e.g.\ Larson 1985, Nagai et al.\ 1998).
Observations of starless cores suggest that they are generally elongated in 
approximately a 2:1 ratio (Myers et al.\ 1991). This provides obvious 
motivation for investigating filamentary models, and is consistent 
with the picture of dense core formation in which a parent molecular cloud 
becomes flattened into a sheet, and fragments into filaments which then 
fragment further into dense cores.

The equilibrium structure of polytropic and isothermal cylinders has been
studied by Ostriker (1964). The density is a function of the radial 
coordinate only: $\rho(r) = \rho_c / (1+r^2/8 H^2)^2$, where 
$H=a/\sqrt{4 \pi G \rho_c}$ is the scale height that is equivalent to the 
scale radius $R_0$ in the Bonnor-Ebert analysis. The density is uniform near
the axis of the cylinder but decays ever more rapidly with increasing radius,
asymptoting to a power law of index $p=4$ for $r \gg H$. Half the mass is
encircled at a radius of $r_m=\sqrt{8} H$.
The filament is supported radially by pressure gradients, but is unstable 
in the direction along its axis. Contraction will proceed initially along 
this direction. For an isothermal filament, the length beyond which it
becomes unstable to fragmentation is roughly four times the half-mass 
radius (Larson 1972, Bastien 1983).

It is interesting to note that the density distribution of the isothermal 
cylinder is a special two dimensional case of the Plummer-like model, with 
$r_m=\sqrt{8} H = R_{\mathrm{flat}}$, $\eta=4$ (the value proposed by 
Whitworth \& Ward-Thompson 2001), and with a physical basis for the 
normalization of the density profile. A cylindrical model can therefore 
reproduce the observational properties of pre-stellar cores that provided 
the motivation for the Plummer-like models. In addition, because the 
isothermal cylinder and the Bonnor-Ebert sphere both represent equilibria 
between self-gravity and gas pressure, the spherically averaged density 
profile of an isothermal cylinder can also mimick closely that of a 
Bonnor-Ebert sphere, in particular a flat inner region with a steeply falling 
envelope (Boss \& Hartmann 2001). The recent success of the Bonnor-Ebert and 
Plummer-like models therefore provides strong encouragement for studying 
this type of model.

\subsection{Fitting Model Parameters and Evaluating Fit Quality}
We evaluate the goodness of fit for these various model density
distributions by calculating a reduced $\chi^{2}$, defined as:
\begin{equation}
\chi^2_\nu=\frac{1}{N-m} \sum{ \left( \frac{\ehks_{i}^{\, \mathrm{NTT}} -
\ehks_{i}^{\, \mathrm{model}}}{\sigma_{i}^{\, \mathrm{NTT}}} \right)^2}
\ ,
\end{equation}
where $m$ is the number of free parameters in the model being fitted,
and the sum extends over a particular subset of $N$ stars taken from
the total number detected in both H \& K$_s$ filters. The values of the
observed color excess $\ehks_{i}^{\, \mathrm{NTT}}$ and the uncertainty
$\sigma_{i}^{\, \mathrm{NTT}}$ are calculated using the properties of
the unreddened stellar background population near B335 (see
Section~\ref{sec:background}), assuming each star to have an intrinsic 
color of $(\overline{\hks})_{\,\mathrm{BG}} = 0.16$, with an uncertainty of 
$\sigma_{\, \mathrm{BG}}= 0.17$:
\begin{eqnarray}
\ehks_{i}^{\, \mathrm{NTT}} & = & (\hks)_{i}^{\, \mathrm{NTT}} - 
(\overline{\hks})_{\, \mathrm{BG}} \nonumber \\ 
& = & (\hks)_{i}^{\, \mathrm{NTT}} - 0.16 \ ,\\[3 mm] 
\sigma_{i}^{\, \mathrm{NTT}} & = & \sqrt{\sigma_i^2 + \sigma_{\, 
\mathrm{BG}}^2} \nonumber \\ 
& = & \sqrt{\sigma_i^2 + 0.17^2} \ , 
\end{eqnarray}
where $\sigma_i$ is the uncertainty in the observed $(\hk)$ color
of a given star. This procedure implicitly assumes that the
photometric errors ($\sigma_i$) and the intrinsic $(\hk)$ colors
both have Gaussian distributions. This is an adequate approximation to 
the actual distribution of background colors, and sufficiently good that
the fitting results are not sensitive to it.

In order to include stars at low impact parameter that provide lower limits
on $\ehks$ (ie stars that are detected at K but not at H), we modify the sum
to include any ``limit stars'' for which the lower limit on $\ehks$ is not 
met. The color excess and uncertainty of these measurements are given by:
$\ehks_{i}^{\, \mathrm{NTT}}=\mathrm{H}_{\mathrm{lim}}-\mathrm{K}_{i}-0.16$,
and $\sigma_{i}^{\, \mathrm{NTT}}=\sqrt{\sigma_i^2 + 0.17^2+ 
\sigma(\mathrm{H}_{\mathrm{lim}})^2}$, with 
$\sigma(\mathrm{H}_{\mathrm{lim}})=0.25$ for 
$\mathrm{H}_{\mathrm{lim}}=18.9$.

Since these models are non-linear in the fitting parameters, 
we analyze the uncertainty in the best-fit model parameters using a 
Monte Carlo technique known as the {\em Bootstrap} method (Press et al.\ 
1992). Table~\ref{tab:fits} summarizes the results from the various
$\chi^2$ analyses and these results are discussed below.

\subsection{Spherically Symmetric Analysis}
The radial variation of the $\ehks$ colors in Figure~3 (South-East excluded)
shows a change in behavior near $83''$ or 0.1~pc (at 250~pc distance), 
apparently the boundary/edge of the dense core and the more extended 
distribution of gas. 
This ``edge'' is smaller than the extent of the core's profile in dust 
emission measured by Visser (2000), but is consistent considering that the 
emission profile is an annular average that includes emission from the 
South-East extension of the dense core. To separate the behavior of the 
core from the more extended and diffuse gas, we restrict the fitting region 
to $r \le 83''$ or $0.1$~pc from the nominal center of the L694--2 core, and 
exclude the wedge of angular size $60^{\circ}$ that is shown in the 
pseudo-image of
Figure~3 as two dashed lines. The column density distributions for the 
various physical models invariably produce profiles that are either flat or 
concave-downwards when viewed on a log-log scale. The fact that the extended
distribution of gas causes the color-excess profile to be concave-upwards 
beyond this edge, necessitates this conservative fitting region to prevent 
the extended extinction from biasing the density structure fit towards 
shallower profiles. This fitting region contains a total of 185 stars with 
measurements at both H \& K$_s$. The origin for the spherically symmetric 
models is left as a free parameter in the following analysis.

Ideally one would like to separate the behavior of the core from the more
extended gas distribution. An obvious approach would be to try to constrain
a composite model that fits for the profile of the core, plus some 
additional constant screen of extinction along the line of sight. This 
would be akin to assuming that the more extended distribution of gas
forms either a uniform level of extinction that covers all lines of sight, 
or perhaps a second envelope in which the L694--2 core is embedded. 
Unfortunately this type of model is in general difficult to constrain. The
variation in colors at a given impact parameter is sufficiently large that
there is too much degeneracy between the model profile and the screen 
extinction at the edge of the globule. Essentially, small variations in the 
screen level can produce very substantial (order of magnitude) variations in
the inferred color excess at the outer edge of the core. In the case of the 
power-law models this makes it impossible to constrain a power-law index and
extinction screen simultaneously. We address this problem by performing the 
fitting analysis for a given model with two different screens of extinction:
a screen of zero (fits marked ``a'' in Table~\ref{tab:fits}), that provides 
a lower limit on the steepness of a given model; and a screen of $\hks=0.19$
(fits marked ``b'') that duplicates the mean extinction level away from the 
core itself (see Section~\ref{sec:background}).

The power-law model that best fits the extinction over the effective range
in radius 0.03~pc$<r<$0.10~pc ($30''<r<83''$) has index $p=2.6 \pm 0.2$,
with $\chi^2_{\nu}=1.114$ (Fit~Ia). The number density normalization of the 
model is $n_{H_2}(0.1$~pc$) = 9.0 \times 10^3$~cm$^{-3}$ (assuming standard 
reddening law and gas-to-dust ratio). The outer radius of the model is not
constrained; the quoted fit is for an outer radius of 0.15~pc ($125''$), but 
any outer radius of $> 0.1$~pc  ($83''$) produces a statistically 
indistinguishable result. The
offset of the central position is $\Delta$R.A.$= -4'' \pm 2''$, 
$\Delta$Dec.$= 4'' \pm 2''$ (corresponds to $x=4''$, $y=4''$ in the pseudo 
image in Figure~3). With an additional $\hks=0.19$ screen of extinction
the fitted slope is much steeper, $p=3.6 \pm 0.3$, and the number density 
normalization at 0.1~pc radius is reduced to 
$n_{H_2}(0.1$~pc$) = 5.5 \times 10^3$~cm$^{-3}$ (Fit~Ib). The fit is improved
($\chi^2_{\nu}=1.063$), and the offset of the central position is unchanged
to within 1~$\sigma$.

For the Bonnor-Ebert models, the column density near the edge of the globule
is not so sensitive to the fitting parameters as in the power law case. For 
these models it is possible to obtain a robust fit for an arbitrary 
extinction screen, but the preferred fit is for an extinction screen of 
zero. In fact for the screen of $\hks=0.19$, the fit fails to constrain a 
model: a Bonnor-Ebert profile cannot be constructed that produces a steep
enough slope over a large enough range in radius in the context of this 
physical scenario. The best fit Bonnor-Ebert model has 
$\xi_{\mathrm{max}}=25 \pm 3$ (Fit~IIa), scale radius 
$R_0=(6.0 \pm 0.5)\times 10^{-3}$~pc ($5.0'' \pm 0.4''$), and physical outer
radius $R_{\mathrm{out}}=0.15 \pm 0.014$~pc ($125'' \pm 12''$).
The center-to-edge density contrast corresponding to the fitted 
$\xi_{\mathrm{max}}$ is $365 \pm 90$, well in excess of the critical value.
The reduced $\chi^2$ for this model is $\chi^2_{\nu}=1.117$, with offset 
central position identical to that for the power-law model. Unlike the power
law models, the Bonnor-Ebert model is based on theoretical argument, and has 
an intrinsic density scale that is related to the physical size of the 
model. For the sound speed of $a=0.20$~km~s$^{-1}$, and distance of 250~pc, 
the central number density of this model is 
$n_{H_2}(0)=3.1 \times 10^5$~cm$^{-3}$, the mass is $M=3.0$~M$_{\odot}$, and
the external pressure is 
$P_{\mathrm{ext}}=1.5 \times 10^{-12}$~dyne~cm$^{-2}$. Note that the density,
mass, and pressure, scale with sound speed and distance as:
$n_{H_2} \propto a^2/R^2$, $M \propto a^2 R$, 
$P_{\mathrm{ext}} \propto a^4/R^2$, respectively. In order to reproduce the 
normalization of the observed color-excess, the Bonnor-Ebert model must be 
scaled by a factor of $\cal{F}=$4.30. In the context of this model, the 
scaling factor must be related to the terms of our adopted conversion 
factors by:
\begin{equation}
\label{eqn:scaling}
\mathcal{F}=\left(\frac{a}{0.20 \, \mathrm{km} \, \mathrm{s}^{-1}} 
\right)^2 \left(\frac{d}{250 \, \mathrm{pc}} \right)^{-1} 
\left(\frac{\mathrm{Gas/Dust}}{2 \times 10^{21} \, \mathrm{cm}^{-2} \,
\mathrm{mag}_V^{-1}} \right)^{-1} \left(\frac{A_V/\ehks}{16.3} \right)^{-1}
\end{equation}
The actual density, mass and pressure will therefore scale differently 
depending on how the factor of 4.3 is distributed amongst these scalings.
We discuss possible implications of this required scaling factor later.
Some subtleties involved in the fitting of Bonnor-Ebert models are 
discussed in the Appendix.

Figure~4 shows the radial dependence of the $\hks$ colors (as in Figure~3)
with the best fitting profiles for the power-law (solid line) and 
Bonnor-Ebert (dashed line) models from Fit~Ia and Fit~IIa respectively. The 
fitting region is marked with a dotted line. The adopted origin in the plot
is based on the results of Fit~IIa (almost identical to that of Fit~Ia). The
two models are indistinguishable over the $26'' < r < 83''$ 
(0.03~pc$<r<$0.10~pc) range in radius where there are stars to fit, as might 
be expected from their almost identical values of $\chi^2_{\nu}$. Only at 
radii $r \lesssim 20''$ (0.024~pc) do the two models begin to diverge 
significantly, due to the fact that the profile of a Bonnor-Ebert sphere 
begins to flatten at roughly $\sim 3$ times the scale radius $R_0$. 

The inferred power-law index of $p \geq 2.6 \pm 0.2$ is far steeper than 
the value of $p=2$ for the singular isothermal sphere. The inside-out 
collapse model is therefore inconsistent with the envelope density structure 
of the L694--2 core, in addition to being inconsistent with the extended 
nature of the infall in this starless source. On this basis, we have 
refrained from fitting this model.

The steep nature of the inferred density profile agrees well with the 
Visser (2001) measurement of $p=2.7$, and with the profiles of other 
contracting starless cores, e.g.\ L1544, for which the dust emission profile
can be reproduced by both a massively super-critical Bonnor-Ebert sphere 
($\xi_{\mathrm{max}} \sim 42$, Evans et al.\ 2001) or a Plummer-like model 
with outer power-law index $\eta=3$ (Lee et al.\ 2003). The case of L1544
demonstrates that a Plummer-like model can be constructed that is essentially
indistinguishable from a super-critical Bonnor-Ebert sphere. The steep
structure of L694--2 inferred from our extinction data is quantitatively
consistent with the Plummer-like model. But the smallest radius
probed by the data is right at the turn-over radius observed in dust 
emission by Visser (2001), and the innermost extinction measurements show no
indication of any flattening in the density profile. We are therefore unable
to constrain the remaining free parameters in the Plummer-like model, 
$R_{\mathrm{flat}}$ and $\rho_{\mathrm{flat}}$. However, unlike the 
Plummer-like model, the turn-over radius for the Bonnor-Ebert sphere is 
intrinsically linked to the density profile in the outerregions based on the
assumption of hydrostatic equilibrium. While the turn-over radius is 
unconstrained in the context of the Plummer-like model, it is constrained in
the context of the Bonnor-Ebert model. The Bonnor-Ebert model 
that successfully matches the data (Fit~IIa) has scale-radius of
$R_0=5''$ ($6\times10^{-3}$~pc), that corresponds to the density profile 
turning over at around $r \simeq 15''$ (0.02~pc). This is around a factor 
of two smaller than the turn-over radius inferred by Visser (2001). However,
this discrepancy
can be explained by Visser's use of an isothermal temperature profile; more
sophisticated modeling of the dust temperature distribution by Evans et al.\
(2001), indicates regions of relatively constant density that are 
systematically smaller (by factors of 0.5--0.9) than are obtained with the
isothermal approximation.

\subsection{Analyzing the Departures from Spherical Symmetry}
The fact that the density profile is flat along the axis of the filament 
essentially allows the gas to support a steeper equilibrium density gradient 
in a direction perpendicular to this axis than can be achieved by a
spherically symmetric model. The steep slope of the fitted power-law model 
that is well in excess of $p=2$ suggests that an isothermal cylinder might 
successfully reproduce the extinction data. Fitting a cylinder that is 
viewed end-on (line of sight parallel to the axis) provides an obvious 
starting point, and by introducing a tilt angle to the analysis permits an 
attempt to model the asymmetry observed in the L694--2 core.

The isothermal cylinder that best fits the same data sample as for the
spherically symmetric model has scale height $H=0.0127 \pm 0.002$~pc or 
$10.5'' \pm 2.0''$, with $\chi^2_{\nu}=1.072$ (Fit~IIIb). 
The central (on-axis) number density of the model is 
$n_{H_2}= 7.7 \times 10^4$~cm$^{-3}$, and in order to reproduce the 
observed normalization of the extinction data with a standard gas-to-dust 
ratio, distance, etc., the cylinder must have length $L=0.6 \pm 0.2$~pc. The 
quality of this fit is comparable to that of the best fitting power-law 
model (Fit~Ib) that includes the same screen of extinction. For the 
cylindrical model, a fit that includes no screen of extinction provides a 
poor description of the data ($\chi^2_{\nu}=1.167$ as compared to 
$\chi^2_{\nu}=1.114$ for a power-law model). In the context of a model
with no additional screen of extinction, the profile of the cylinder can
only reproduce the overall slope of the extinction data with a large value
of the scale height, but this introduces a curvature to the profile that
is not found in the data.

The success of the screened cylindrical model suggests an intriguing 
physical description of the L694--2 core. The core may represent a region
of over-dense material that has condensed out of the extended filament of
dense gas that is seen in the DSS image (Figure~1). Since the L694--2 core
itself is clearly elongated along the South-East direction, the obvious next
step in the model-fitting process is to try to constrain an isothermal
cylinder of length $L$ that is tilted by an angle $\phi$ to the line of 
sight (in such a direction that its projected axis $L \sin(\phi)$ is along 
a South-East direction), and is
embedded in a a more extended distribution of gas that produces an extra 
screen of extinction (i.e.\ as for Fit~IIIb). The fitting region for this 
tilted model should be modified to include stars along the South-East 
direction, to test whether the tilted cylinder can successfully reproduce 
the observed asymmetry of the L694--2 core. For a dataset of all stars 
within $r < 0.1$~pc ($83''$) of the N$_2$H$^+$ peak, the best fitting model 
has scale height $H=0.0164 \pm 0.002$~pc or $13.5'' \pm 1.5''$, with
$\chi^2_{\nu}=1.189$ (Fit~IVb). The central (on-axis) number density is
$n_{H_2} = 4.7 \times 10^4$~cm$^{-3}$. The top of the tilted cylinder is
offset from the mm-peak by 
$\Delta$R.A.$= -30'' \pm 3''$, $\Delta$Dec.$= 30'' \pm 3''$, 
although the top of the cylinder does not represent the peak column density
of the model. The length and tilt angle of the cylinder are not 
constrained individually if one allows for uncertainties in the gas-to-dust
ratio, distance etc. However, the projected length of the cylinder can be
constrained: $L \sin(\phi) = 0.14 \pm 0.02$~pc or $117'' \pm 17''$, with the 
scaling given by $\mathcal{F}={}$ $(1.1 \pm 0.1) (0.5 \, \mathrm{pc}/L)$. 
Figure~5 shows schematic diagrams of the 3-D geometry of this best fit 
tilted cylindrical model.

This model is remarkably successful at describing the extinction data from 
all parts of the L694--2 core. The dataset is only increased by 14
stars by the inclusion of the South-East region, and the $\chi^2_{\nu}$ is
increased by around 0.12 from the best-fitting spherically symmetric models.
Yet this represents the model (on average) reproducing the high extinction 
of all of these 14 stars to within 1~$\sigma$ in their measurement 
uncertainties. Moreover, if one attempts to fit a spherically symmetric model
to this same dataset, the fit is drastically worse: $\chi^2_{\nu}=2.14$ for
a best-fit power law model with an extinction screen. This further heightens
the degree of success with which a tilted cylinder can describe both the 
radial and azimuthal variation of the color-excess. A pseudo-image and
the radial color dependence of a best-fit tilted cylinder model with
length of $L=0.5$~pc is shown in Figure~6.

\subsection{Discussion}

The models that have been investigated have highly idealized geometries, and 
represent extreme pictures of the L694--2 core. Nevertheless, the success of 
these models is noteworthy, and provides a useful basis for interpreting
the extinction observations, and drawing conclusions on the physical 
structure of the dense core. Figure~7 shows a plot of the molecular hydrogen 
number density profiles of the best fit models from Table~1. The various
models have similar profiles in the region where color-excess measurements
can be obtained ($r \geq 30''$, 0.035~pc or 7500~AU). In the inner regions 
the profiles differ significantly; future observations of L694--2 in dust
emission made with interferometers may therefore distinguish 
between the various interpretations of density structure.

The Bonnor-Ebert model is a hydrostatic equilibrium structure, so a fitted 
value of $\xi_{\mathrm{max}}=25$ that is so far in excess of the critically 
stable value presents an obvious problem with a Bonnor-Ebert description 
of L694--2. The analysis does indicate that the distribution of material in 
L694--2 is very unstable to gravitational collapse, a conclusion that is 
consistent with the strong infall signatures observed in molecular spectral 
lines. Including an additional screen of extinction, as might be expected 
since the L694--2 core is clearly embedded in a more extended distribution 
of gas, causes the fit for a Bonnor-Ebert model to fail.
Essentially this model is unable to reproduce the steep slope of the inferred
density profile if any additional extinction is present.

The Bonnor-Ebert model provides a description of the dynamical state of the 
core that is absent from the simple power law models. Unfortunately, the 
normalization of these models differs substantially from what is observed in 
the L694--2 core ($\mathcal{F}={}$$4.3$). Remarkably, the difference in the 
normalization between observation and theory is quite similar to that found 
in the study of B335, a core that is likely associated with the same parent
molecular cloud as L694--2 ($\mathcal{F}={}$$5.2$ for an inside-out collapse 
model, $\mathcal{F}={}$$3.3$ for a Bonnor-Ebert sphere).

The origin of this required scaling presented a difficult problem in the
B335 study, since it is hard to attribute such a large discrepancy to any one
stage in the conversion from density to color-excess. However, the present
study suggests an intriguing explanation for the surprisingly high 
extinctions observed in these cores, namely that the discrepancy is simply
a result of the cores having larger masses than predicted by the spherically
symmetric theory, perhaps with B335 representing simply a later stage to 
which the L694--2 core will eventually evolve. The L694--2 core has formed
out of a filament of dense gas, and our modeling suggests that it has 
prolate structure, with the major axis lying close enough to the 
line-of-sight that much of the mass is missed by analysis that assumes a 
symmetry between the profile of the core in the plane of the sky, and the
profile of the core along the line-of-sight. The infalling motions inferred
from molecular spectral lines by Lee et al.\ (2001) represent motions along
the line of sight. If the motions arise in the core and not the
surrounding envelope of material, then the strength and orientation of these 
motions is consistent with our hypothesis, since a prolate core is most 
unstable to collapse along its major axis.

The relation of the L694--2 dense core to the associated lower-density 
filament seen in the DSS image is interesting: the core appears to
be elongated along the line of sight, while the filament appears to be
elongated in the plane of the sky. A simple explanation for the origin of 
such this system is as follows. Suppose a low-density filamentary cloud is 
extended in the plane of the sky. The cloud has some angular momentum and 
is very slowly rotating about an axis perpendicular to its long 
direction.  A small elongated region of the cloud condenses while 
conserving its angular momentum, therefore turning faster than the 
rest of the cloud. The result is a short dense filament whose major 
axis is rotated with respect to the long axis of its lower-density 
parent filament, which has become ``kinked''. In the L694--2 case, the 
short dense segment seems to lie mostly along the line of sight.

Theoretical argument and numerical simulation both suggest that an 
un-magnetized filament will be unstable to fragmentation if its length 
exceeds about four times its half-mass radius. The half-mass radius for the 
fitted tilted cylindrical model of the L694--2 core is $0.05$~pc ($40''$), 
which suggests a maximum stable length of only about $0.2$~pc. 
This means that for the standard conversion factors, the length of the 
fitted cylinder is such that it should fragment into two or more pieces at 
some stage in the future (or possibly should have already done so), unless 
some source of support such as a magnetic field acts to prevent this process.
If fragmentation of the L694--2 core does not occur, then future evolution
should produce a more spherically symmetric core, with a mass that is larger
than would have otherwise been expected. Subsequent collapse of this core
might resemble that observed in B335.

This hypothesis requires a special orientation of the magnetic field in
the L694--2 and B335 cores, both to support the cores, but also to prevent
the support from affecting the observed molecular line widths. Although this 
may seem contrived, the orientation of B335 is well known to be special, with
the outflow lying within $10^{\circ}$ of the plane of the sky 
(Hirano et al.\ 1988). Similarly, our modeling of the color excess near 
L694--2 suggests a prolate core oriented close to the line-of-sight 
($20^{\circ}$ for $L=0.4$~pc). Observations of optical and infrared 
polarization of stars that are embedded in, or background to filamentary 
molecular clouds show that the direction of the magnetic field is often
related to the direction of the axis of the filament; in some cases the two 
are parallel (e.g.\ Ophiuchus, Goodman et al.\ 1990), while in others the 
two are perpendicular (e.g.\ Taurus, Heyer et al.\ 1987). Both cases can be
explained by recent modeling of the production of gas filaments
by the fragmentation of magnetized sheets (Nagai et al.\ 1998). In these
models, the magnetic field delays the growth of perturbations along a 
particular direction, causing the production of filamentary structures.
The field provides support to the filament, and increases the critical
(fragmentation) length of the filament from the un-magnetized case. If the
field in L694--2 is oriented in the plane-of-the-sky, perpendicular to the
major axis of the core, then the support provided along the major-axis would 
be maximized. Since the turbulent velocity in L694--2 is low 
($a_{\mathrm{turb}}\simeq 0.09$~km~s$^{-1}$ compared to the thermal speed of 
$a_{\mathrm{therm}}\simeq 0.18$~km~s$^{-1}$), Alfven waves should not be 
strongly driven, and not add 
appreciably to measured line-widths. Such a field configuration might 
therefore inhibit fragmentation but remain essentially undetectable in
observations of molecular spectral lines.

To affect the fragmentation scale of L694--2, the field strength would need 
to be such that the magnetic pressure ($P_M = B^2/8 \pi$) is greater than or
comparable to the gas pressure at the axis of the filament 
($P_g= \rho_0 a^2$). For the standard conversion factors, this suggests:
\begin{equation}
B \gtrsim \mathrm{few} \, \times 10 \mathrm{~}\mu\mathrm{G}
\end{equation}
Such a minimum field strength is large by typical standards of the 
Interstellar Medium. For instance, Crutcher \& Trowland (2000) measured a 
line of sight magnetic field of $B \sim 11$~$\mu$G in starless core L1544 
using the Zeeman effect in lines of OH, although any non-zero inclination 
angle will result in the true field being stronger. Although the L1544 field 
is already much higher than the typical upper limits found for other dark 
clouds (Crutcher et al.\ 1993), a field of this required strength in the 
L694--2 core is certainly conceivable. Moreover, if any of the required
scaling can be attributed to systematic error in the conversion factors,
then the minimum field strength will be lowered. For instance, a gas-to-dust
ratio in the L694--2/B335 region that is half the ``standard'' value in the
ISM (a variation that is typical in the ISM), then the inferred length of
the tilted cylinder (Fit~IVb) will be reduced by roughly the same factor of
two, and the cylinder would be marginally stable to fragmentation even in
the absence of a magnetic field.

\section{Summary}
We present a near-infrared extinction study of the contracting starless
core L694--2 using observations made with the ESO NTT. In summary:

\begin{enumerate}
\item The $5' \times 5'$ image of L694--2 shows a dramatic fall off in the 
number of stars detected toward the location of peak millimeter-wave 
(N$^2$H$^+$) emission, where the extinction increases because of the central
concentration of dense core material. The image contains a total of 1451 
stars detected in both H \& K$_s$ bands, with 199 with impact parameters 
within $83''$ (0.1~pc), the innermost of which is at $28''$ (0.034~pc)
from the nominal peak. The photometry shows a steep gradient in the $\hks$ 
colors towards the center. The radial profile of the color excess 
demonstrates a change in behavior at a radius of around $83''$ (0.1~pc): the 
profile 
flattens, apparently associated with the boundary/edge of the dense core and
a more extended distribution of gas in which the core is embedded. In 
addition the core exhibits departures from spherically symmetry, with the 
regions of highest color excess extending to the South-East, and somewhat
towards the North, following the shape of L694--2 visible at low extinction 
levels in the DSS image. These asymmetries are apparently 
connected to the filamentary structure of the extended gas in the region. 

\item We compare a series of models of dense core structure to the extinction
data, including spherically symmetric power-laws, and Bonnor-Ebert spheres. 
Based on the radial profile of the color excess, only stars 
within $83''$ (0.1~pc) of the peak are included in the fit to prevent the 
extended extinction from biasing the fitting results towards models with 
shallower slopes. In addition, for the spherically symmetric analysis, we 
exclude stars that fall within a small wedge to the South-East ($60^{\circ}$
wide), to avoid biasing the fits with the extinction component that extends
in this direction. It is not possible to fully separate
the contribution to the color excess from the core from that due to 
additional extended material. We therefore perform two fits for each
type of model, one assuming extinction only from a model core, and the 
second assuming extinction from a core embedded in a screen of thickness 
$\hks=0.19$, chosen to reproduce the colors of stars at the edge of the 
image where the color excess is smallest (i.e.\ $> 2\farcm 5$ from the 
center). The best fit single power law model has index $p=2.6 \pm 0.2$ 
(1~$\sigma$), steeper than the value of $p=2$ for an isothermal sphere. 
An unobscured Bonnor-Ebert sphere provides an indistinguishable fit to the
data; the dimensionless outer radius is $\xi_{\mathrm{max}}=25 \pm 3$. 
Including an additional uniform extinction component increases the steepness
of the inferred core profiles. The best fit power law index increases to 
$p=3.7 \pm 0.3$, while a Bonnor-Ebert model cannot be constructed that 
produces a steep enough slope over a large enough range in radius. 
The unobscured Bonnor-Ebert model must be scaled by a factor 
$\cal{F}={}$4.3 to match the observed color excess.
Remarkably, this difference in normalization between observation and 
theory is similar to that found in the extinction study of B335, which is
likely associated with the same molecular cloud.

\item The inferred power law index $p$ suggests a cylindrical model for
the density distribution, which can support a large density gradient
perpendicular to its cylindrical axis. An embedded cylinder viewed along the 
axis with scale height $H=0.0127 \pm 0.002$~pc ($10.5'' \pm 2.0''$) 
provides an equally good fit as a spherical power law model. A tilted 
cylinder can also
reproduce the asymmetry of the L694--2 core, matching the color excess of 
stars in the South-East wedge to within 1~$\sigma$ on average. This tilted 
cylinder has projected length $L \sin{\phi} = 0.14 \pm 0.02$~pc 
($117 \pm 17''$), with scale height $H=0.0164 \pm 0.002$~pc  
($13.5'' \pm 1.5''$). For standard values of the conversion factors, the 
model must be scaled by a factor of 
$\cal{F}={}$$(1.1 \pm 0.1)(0.5\mathrm{~pc}/L)$ to match the observed 
color excess.

\item The cylindrical model provides an intriguing explanation for the 
surprisingly high extinctions observed in the L694--2 core, and perhaps also
the B335 core: the discrepancy may result from larger masses than indicated
by spherically symmetric theory, perhaps with B335 representing a later 
stage to which the L694--2 core will eventually evolve. The large scale view
suggests that L694--2 has formed out of an extended filament of dense gas, 
and our modeling suggests a prolate structure, with the major axis lying 
close enough to the line-of-sight that much of the mass is missed by 
assuming symmetry between the profile of the core in the plane of the sky, 
and the profile of the core along the line-of-sight. The inward motions 
in L694--2 inferred from molecular spectral lines, which represent motions 
along the line of sight, might reflect the fact that a prolate core is most
unstable to collapse along the major axis. If the cylinder is
magnetized, with static field $B_0 \gtrsim \mathrm{few} \times 10$~$\mu$G
in the plane of the sky, then fragmentation of this cylinder could be
prevented. The low turbulent velocity measured in L694--2 should prevent
Alfven waves from contributing to the observed widths of molecular spectral 
lines, and the support would remain essentially undetectable.
Moreover, if the gas-to-dust ratio in this region is lower than the 
``standard'' value, then fragmentation might be avoided at a lower field 
strength.

\item Future observations of dust in emission made with interferometers
will measure the innermost structure of the L694--2 core that can not be 
probed by dust extinction. In particular, the dust emission structure will
provide further tests of the cylindrical model hypothesis.
\end{enumerate}

\acknowledgements
DWAH is indebted to Tracy Huard for his help during the data reduction
stages. We also thank Lee Hartmann for useful suggestions regarding
departures from spherical symmetry. This publication makes use of data 
products from the Two Micron All Sky Survey, which is a joint project of 
the University of Massachusetts and 
the Infrared Processing and Analysis Center/California Institute of 
Technology, funded by the National Aeronautics and Space Administration 
and the National Science Foundation. CJL acknowledges support
from the NASA Origins program, Grant NAG-5-9520.

\appendix

\section{Subtleties in Fitting Bonnor-Ebert Models}
It is interesting that both the scale-radius and outer-radius for the 
fitted L694--2 Bonnor-Ebert model (Fit~IIa) are better constrained than the 
dimensionless radius of the sphere, when the three parameters are 
intrinsically linked by the relation: 
$R_{\mathrm{out}}= \xi_{\mathrm{max}} R_0$. 
The Bonnor-Ebert solution is a dimensionless curve that 
extends from the origin to a cutoff at the dimensionless outer radius 
$\xi_{\mathrm{max}}$. Scale is added by including a physical size for the 
curve (either the scale radius or the outer radius). A given density
profile over a fixed range in radius can therefore be reproduced by a model
with a larger physical size and a larger $\xi_{\mathrm{max}}$, thereby 
assuring that we are still looking at the same region of the dimensionless 
curve. Such a model will have the same value of the scale-radius $R_0$.

In the model fitting procedure, what is actually observed is the integrated 
profile of the globule, and this is altered slightly due to the contribution
from the extra material at the edge of the globule, which produces a more 
shallow profile. However, except for right at the edge of the globule, the 
effect of the extra material on the shape of the profile can be offset by 
reducing slightly the scale-radius of the model (or increasing 
$\xi_{\mathrm{max}}$ a little more), which raises the degree of central 
concentration and steepens the profile a little. It is the change to the 
profile near to the outer radius from this extra material that allows the 
outer radius and $\xi_{\mathrm{max}}$ to be constrained by measurements of 
an extinction profile.

To summarize, despite the fact that the family of Bonnor-Ebert spheres are 
characterized by the parameter $\xi_{\mathrm{max}}$, it is the parameter 
$R_0$, the scale-radius, that is the more robust result of a fit to an 
observed profile. The outer radius of a globule has some inherent and 
non-zero uncertainty, due to possible confusion with extended structure, 
nearby sources or interaction with the ambient medium or interstellar 
radiation field (ISRF). The effect is that the physical curvature is 
inevitably better defined than the dimensionless curvature, since the latter
suffers the additional contribution from the uncertainty in the radius (in 
order to make the observed profile dimensionless, the radii must be divided 
by the outer radius). 

When Bonnor-Ebert models were fitted to B68 (Alves et al.\ 2001) and B335
(Harvey et al.\ 2001), fixed outer radii were assumed. The uncertainty in
the assumed outer radii were not included in the final assessment of the 
fitted profiles, which will lead to underestimated errors in 
$\xi_{\mathrm{max}}$. The systematic uncertainty in the assumed radius will 
also lead to systematic uncertainties in $\xi_{\mathrm{max}}$, which were
not accounted for in these studies. In both studies the radial variation
of the $\hk$ colors shows that the adopted outer radii represent essentially
lower limits to the true outer edge of the globule, in the B68 case because
the edge is assumed to occur at a visual extinction of $A_V=1$, and in the
B335 case because the edge is chosen to coincide with a low-power contour
of molecular gas emission. Allowing the radius of the globule to vary as an
additional parameter will therefore lead to a larger fitted value of 
$\xi_{\mathrm{max}}$, even for a small increase in the outer radius. The
uncertainty in $\xi_{\mathrm{max}}$ will also increase. In the case of 
B335, the quoted $\xi_{\mathrm{max}}$ is so large (and already uncertain at a
level that is similar to the uncertainty in the outer radius) that
realistic changes in the outer radius cannot lead to a qualitative change
in the picture --- the fitted $\xi_{\mathrm{max}}$ will increase even 
further from the stable value, and although the fitted $R_0$ may decrease 
somewhat, the model would remain heavily resolved by the IRAM PdBI, and
therefore be unable to reproduce the observed mm-wave dust emission (Harvey
et al.\ 2003). 

For B68 however, the conclusions are less clear. The quoted value of 
$\xi_{\mathrm{max}}=6.9 \pm 0.2$ leads to the interpretation that the 
globule is almost perfectly perched at the threshold for stability. A small 
change in the inferred $\xi_{\mathrm{max}}$ might alter this conclusion. 
We therefore apply our more detailed model-fitting procedure to the Alves et
al.\ (2001) extinction data. Unlike the L694-2 data, the B68 data are
azimuthally averaged in bins of width $3''$ centered on the position of
peak column density. The small column density of the B68 globule essentially
allows the profile to be measured at all radii, from $10''$ out to $130''$
where no extinction can be discerned.

A fit to the data within a conservatively chosen region $r \leq 100''$, with
outer radius and $\xi_{\mathrm{max}}$ as free parameters returns essentially 
the Alves et al.\ result: $\xi_{\mathrm{max}}=6.9 \pm 0.2$, 
$R_{\mathrm{out}}=106'' \pm 1''$, with $\chi^2_{\nu}=1.43$. However, with a
different fitting region, the inferred model parameters are slightly 
changed. The value of $\chi^2_{\nu}$ provides a useful tool to 
select the appropriate fitting region. As the region is extended from $100''$
towards the edge of the globule, the minimum in $\chi^2_{\nu}$ remains fairly
constant initially, but then jumps to 2.1 for $r \leq 109''$, and continues
to increase, reaching a value of almost 5 for a fitting region of 
$r \leq 130''$ --- the largest radius at which any extinction was 
measurable. This behavior is caused by the profile beginning to depart from
the Bonnor-Ebert profile at the edge of the globule, perhaps due to 
interaction with the ISRF. We therefore suggest a fitting region of 
$r < 109''$ is most appropriate (included data point at largest radius is at
$r=106''$). In particular this choice happens to coincide with the inferred 
outer radius of the Bonnor-Ebert model, as well as the radius where the 
visual extinction reaches unity. The best fit model 
parameters for this region are $\xi_{\mathrm{max}}=7.2 \pm 0.2$, 
$R_{\mathrm{out}}=108 \pm 1''$, with $\chi^2_{\nu}=1.55$. The increase in
the minimum $\chi^2_{\nu}$ in moving to this larger fitting region is 
justifiable, since it is equivalent to that for a model where the parameters
differ by only 1/2~$\sigma$ from their optimal values. That the Bonnor-Ebert 
model is so tightly constrained results from the well defined edge 
to the B68 globule, in stark contrast to the extended structure associated 
with L694--2. However, the best-fit model with 
$\xi_{\mathrm{max}}=7.2 \pm 0.2$ represents a slightly more unstable 
configuration than the value calculated by Alves et al.\ (2001).

The physical properties of this model differ at most only slightly from 
those of the Alves et al.\ model. The inferred mass of the globule is
$M=2.2$~M$_{\odot}$ for a distance of 125~pc, and the standard gas-to-dust 
ratio. The inferred central number density is increased by roughly 2.5\%, 
but to the level of precision in the factors required for the conversion
it is essentially unchanged $n_{H_2}= 4.0 \times 10^5$~cm$^{-3}$. The 
external pressure is 6\% smaller, 
$P_{ext}=2.3 \times 10^{-11}$~dyne~cm$^{-2}$, for a kinetic temperature of
16~K. Hotzel, Harjo \& Juvela (2002) have proposed a smaller distance to B68
of 85~pc and a lower kinetic temperature of $10 \pm 1.5$~K. These changes
lead to a smaller inferred mass of the globule, $M=1.5$~M$_{\odot}$, a
higher central number density $n_{H_2} \simeq 5.4 \times 10^5$~cm$^{-3}$, 
and a smaller external pressure $P_{ext}=2.0 \times 10^{-11}$~dyne~cm$^{-2}$.

\clearpage
\begin{deluxetable}{llllll}
\tablenum{1}
\tabletypesize{\scriptsize}
\tablewidth{0pt}
\tablecolumns{6}
\tablecaption{Summary of the $\chi^2$ Analyses Data \label{tab:fits}}
\tablehead{ \colhead{Fit} & \colhead{Model Density Distribution} 
& \colhead{Fit Region} & \colhead{Stars} & \colhead{Fitted Model 
Parameter(s)} & \colhead{$\chi^2_\nu$}}
\rotate
\startdata
Ia & Power Law: $\rho \propto r^{-p}$ & $r<83''$; No ``wedge'' & 185 &
$p=2.6 \pm 0.2$ & 1.114 \\
Ib & Power Law & $r<83''$; No ``wedge'' & 185 & 
$p=3.7 \pm 0.3$ & 1.063 \\ 
IIa & Scaled Bonnor-Ebert $\rho \propto \mathcal{F} \rho_{BE}$ & $r<83''$; 
No ``wedge'' & 185 & $\xi_{\mathrm{max}}=25 \pm 3$; $R_{out}=0.15 \pm 
0.014$~pc; $\mathcal{F}={}$ $4.3 \pm 0.3$ & 1.117\\
IIIb & Cylinder (end-on) & $r<83''$; No ``wedge'' & 185 & 
$H=0.0127 \pm 0.002$~pc; $L=0.6 \pm 0.2$~pc & 1.072\\
IVb & Scaled Cylinder (tilted) & $r<83''$ & 199 & $H=0.0164\pm 0.002$~pc;
$L \sin(\phi) = 0.14\pm 0.02$~pc; 
$\mathcal{F}={}$ $(1.1 \pm 0.1) (0.5 \, \mathrm{pc}/L)$ & 1.190
\enddata
\tablecomments{For each Fit, we consider only stars within 0.1~pc or $83''$ of
the nominal center of L694--2, to avoid contamination from the more extended 
distribution of gas. For Fits Ia--IIIb, stars that occupy a wedge of angular
extent $60^{\circ}$ centered along the South-East direction have not been 
considered in order to eliminate the elongation of the core in this direction
from affecting the fit of these azimuthally symmetric models. Fits marked 
``\#b'' include an extra screen of extinction of $\hks=0.19$. }
\end{deluxetable}

\clearpage
\begin{figure}
\figurenum{1}
\epsscale{0.90}
\plotone{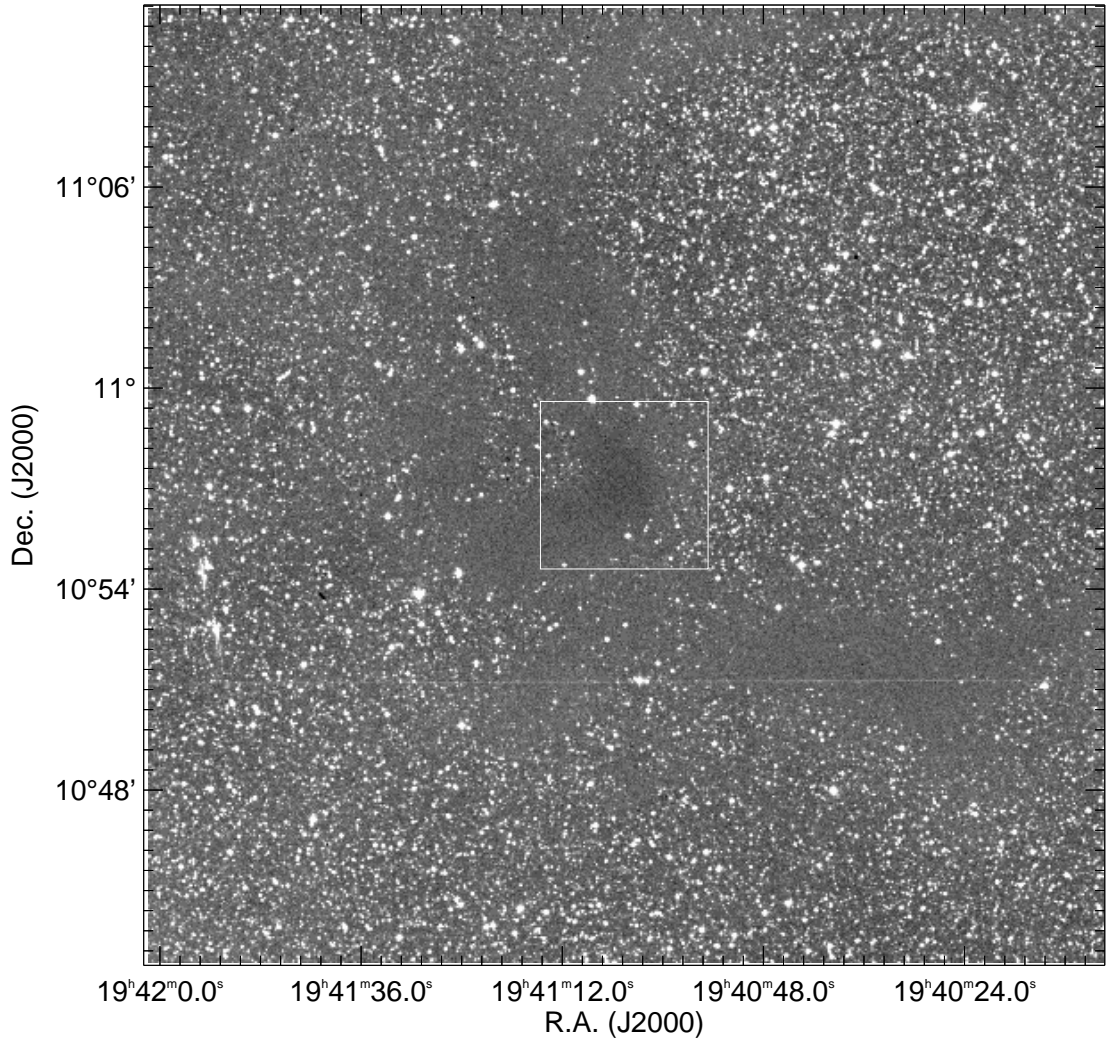}
\caption{Digital Sky Survey POSS II image (Red plate) of the 
region containing L694--2. The field observed with the NTT is shown as 
a white box. The core appears to be a dense knot in an extended and 
filamentary distribution of molecular gas.}
\end{figure}

\clearpage
\begin{figure}
\figurenum{2}
\epsscale{0.60}
\plotone{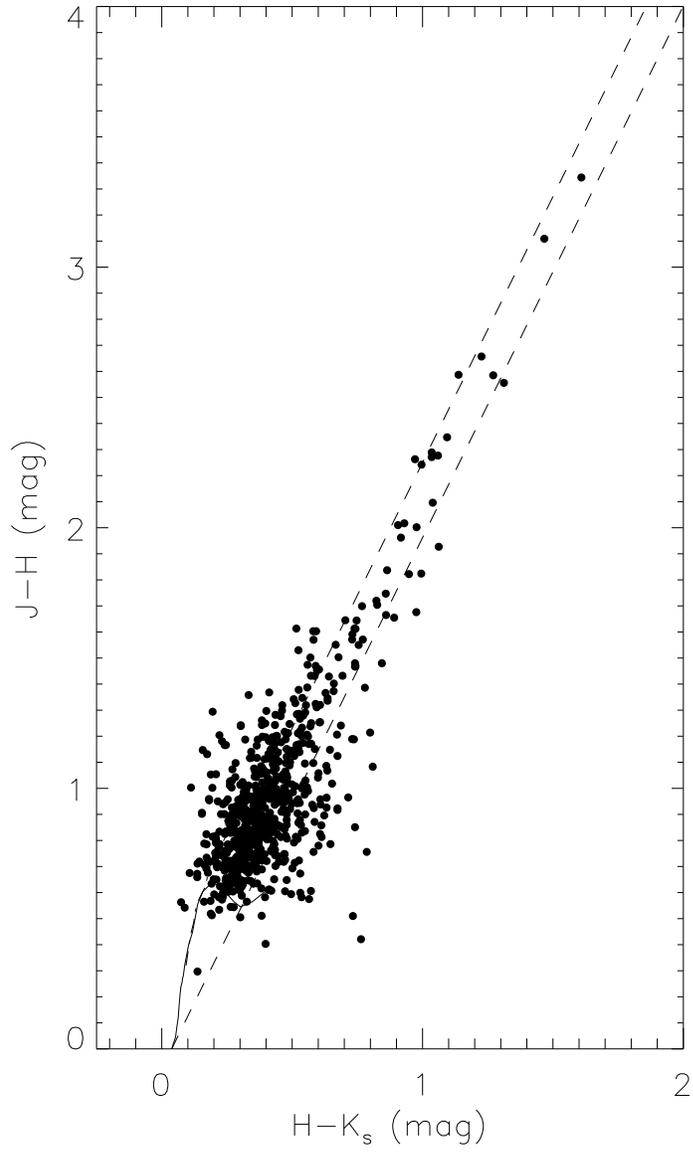}
\caption{JHK$_s$ color-color diagram for the stars near L694--2 
with signal-to-noise $>10$. See text for discussion.}
\end{figure}

\clearpage
\begin{figure}
\figurenum{3}
\epsscale{1.00}
\plotone{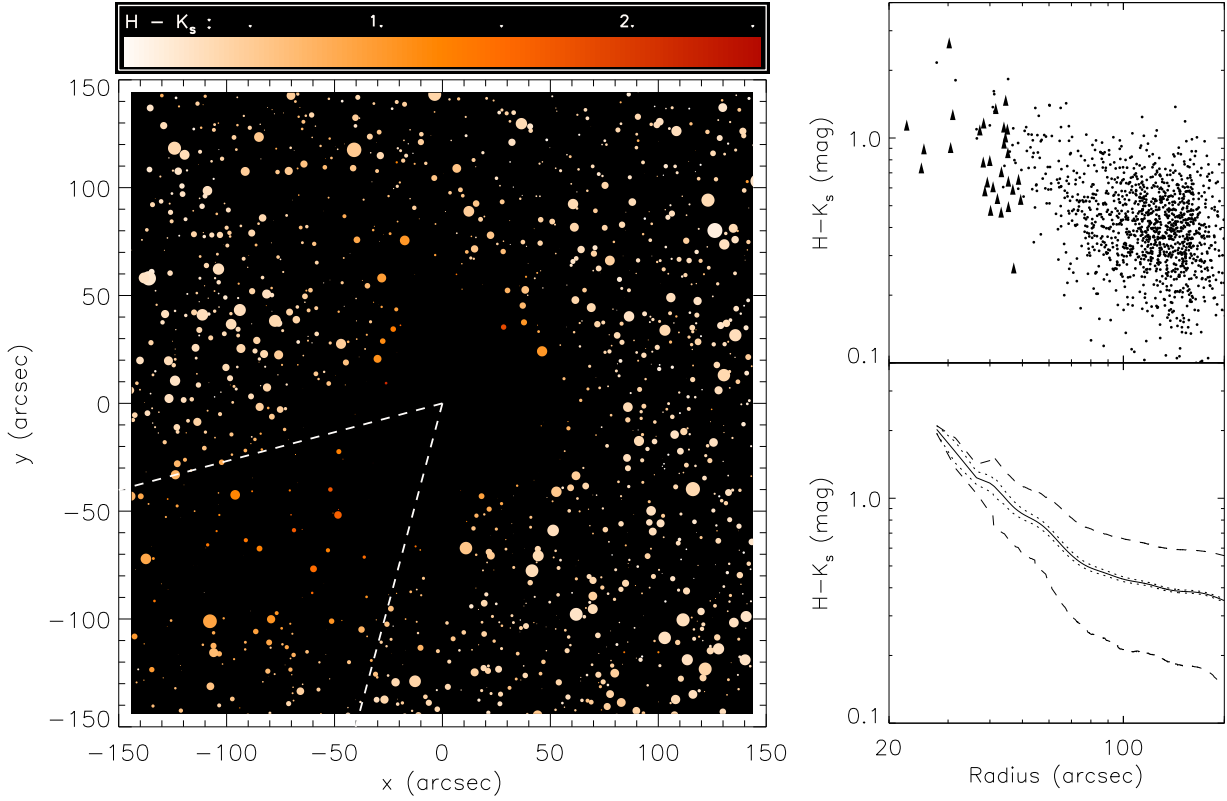}
\caption{Left: Pseudoimage of the NTT/SofI observations; the axes 
are spatial, H magnitude determines a star's ``size'', and $\hks$ determines
its color. The image contains 1451 stars detected in both filters. The 
dashed line indicates a bar-like structure of extinction that extends a 
significant distance from the core. Upper right: $\hks$ color against radius 
out to $200''$ from the nominal center of L694--2. There is a steep gradient 
toward this location. Stars within the dashed region in the pseudoimage are 
omitted from the plot. Photometric error bars are not shown as they cause 
confusion with such a large number of stars. Lower limits on $\hks$ are
marked as upward-pointing triangles. Lower right: smoothed $\hks$ 
color against radius out to $200''$ from the nominal center of L694--2 
(solid line), with $\pm$ one standard deviation (dashed line) and $\pm$ one 
standard error (dotted line). See text for further explanation.}
\end{figure}

\clearpage
\begin{figure}
\figurenum{4}
\epsscale{0.60}
\plotone{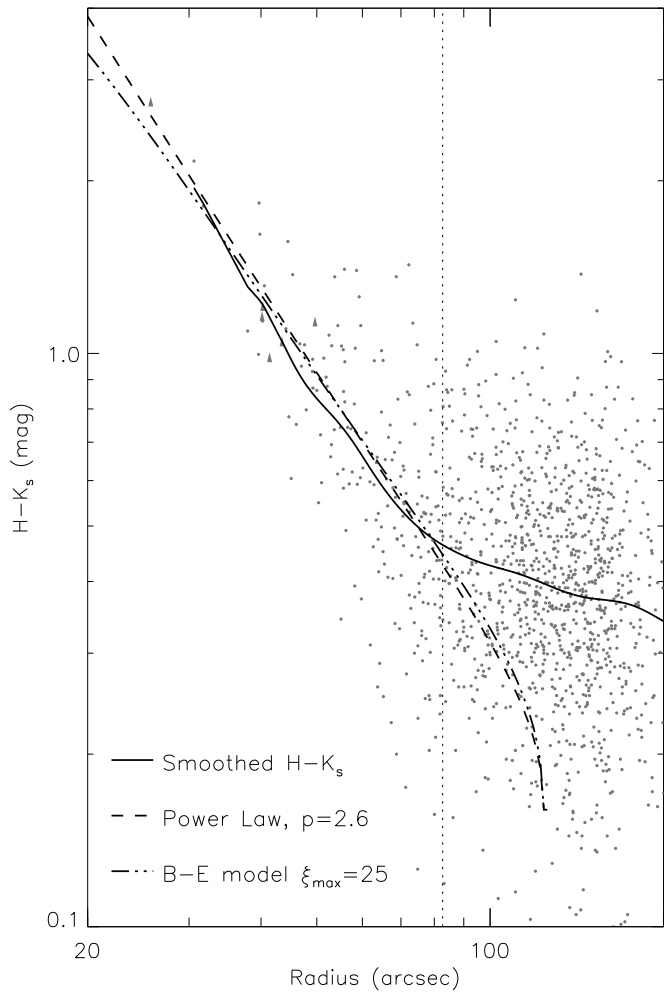}
\caption{$\hks$ color against radius as in the upper-right panel of 
Figure~3. In this plot the origin is defined by the fitted position of the 
center, as opposed to the N$_2$H$^+$ peak used in Figure~3. This center 
adopted in this plot essentially minimizes the dispersion in the colors at 
a given radius. Stars that provided lower limits on $\hks$ that were weak
have been omitted from the plot, for the sake of clarity.
Overplotted are the profiles for the best-fitting power law model 
(Fit~A, dashed line), and Bonnor-Ebert sphere (Fit~B, dash-dotted line), as 
well as the Gaussian smoothed profile (solid line). The outer radius of the 
fitting region is also marked (dotted line). The two models are 
indistinguishable over the range in radius where there are stars to fit; 
only at radii $r \lesssim 20''$ (0.024~pc) do they diverge significantly.} 
\end{figure}

\clearpage
\begin{figure}
\figurenum{5}
\epsscale{1.00}
\plotone{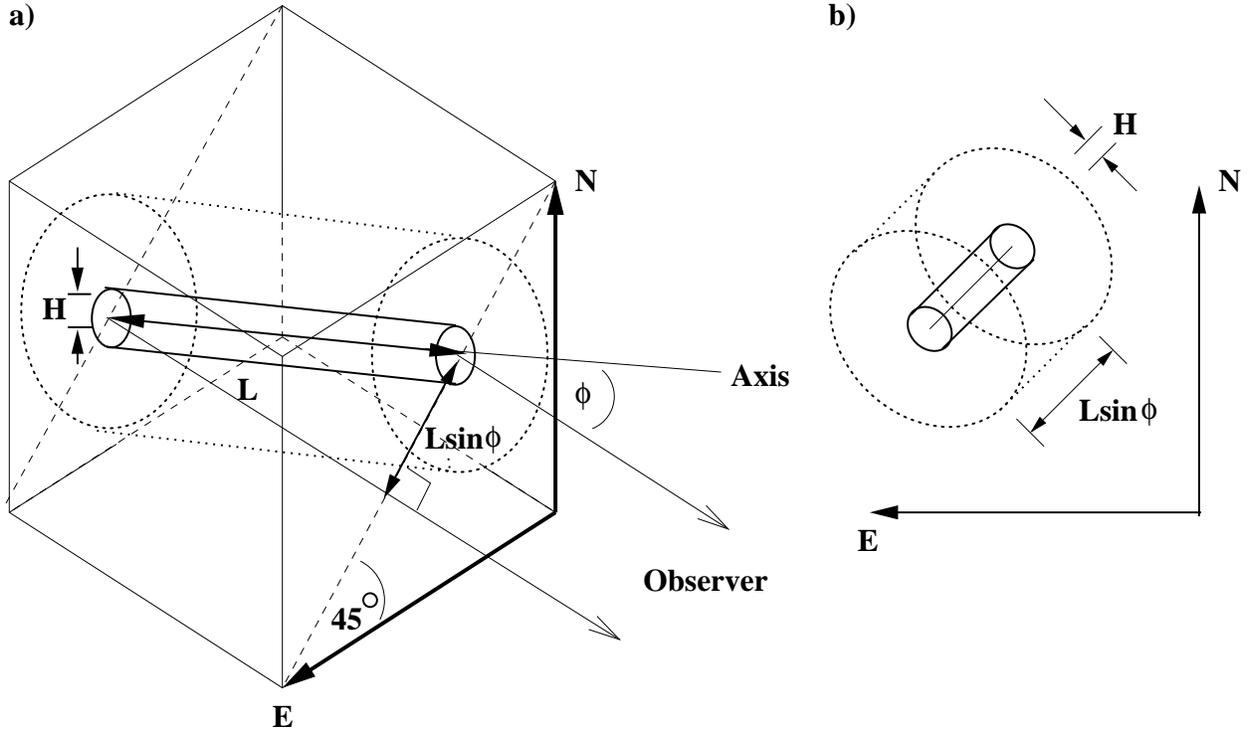}
\caption{(a): Schematic diagram illustrating the 3-D geometry of the tilted
cylindrical model of the L694--2 dense core. (b): the model as projected 
onto the plane of the sky. The best-fit model has scale height 
$H=0.0164 \pm 0.002$~pc ($13.5'' \pm 1.5''$), and projected length 
$L \sin{\phi} = 0.14 \pm 0.02$~pc ($117'' \pm 17''$). The dotted outer 
contour represents where the model column density falls to the level of
the extended distribution of gas in which the core is embedded, a radius of
$\sim 0.1$~pc or $83''$. For the standard conversion factors (gas-to-dust 
ratio, distance, etc.), the model must be scaled by a factor of 
$\cal{F}={}$$(1.1 \pm 0.1)(0.5\mathrm{~pc}/L)$ to match the observed color 
excess. If one allows for the uncertainties in the conversion factors (i.e.\
permitting $\cal{F}{}$$\neq 1$) then the length and tilt of the cylinder 
cannot be constrained individually.}
\end{figure}

\clearpage
\begin{figure}
\figurenum{6}
\epsscale{1.00}
\plotone{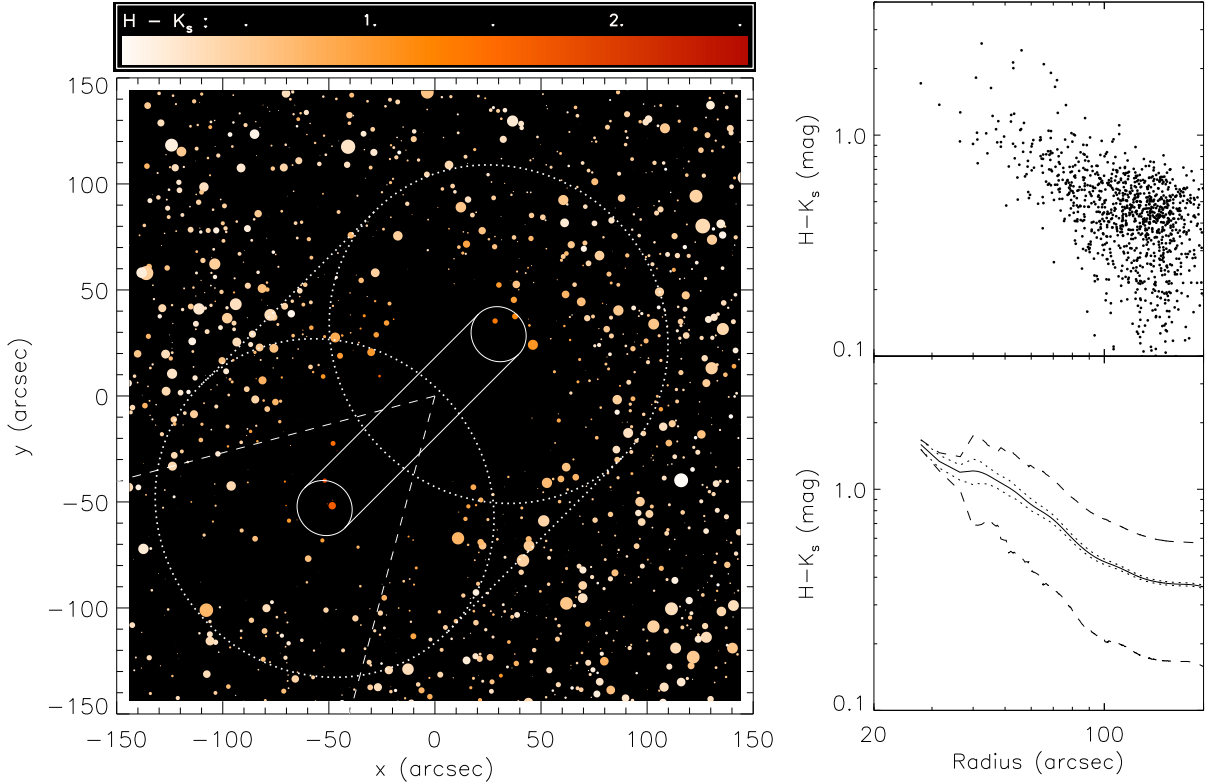}
\caption{Left: Pseudo image of a best-fitting tilted cylindrical 
model (see text for description). The schematic diagram of the model from 
Figure~5b is superposed on the plot. Upper right: $\hks$ color against 
radius out to $200''$ (0.24~pc) from the center. Stars within the dashed 
region in the 
pseudoimage are omitted from the plot. Lower right: smoothed $\hks$ color 
against radius out to $200''$ from the nominal center of L694--2 (solid 
line), with $\pm$ one standard deviation (dashed line) and $\pm$ one 
standard error (dotted line). The success of the model can be seen by 
comparison with the actual color excess data in Figure~3. Note that the 
dispersion in the $\hks$ colors at a given radius is due to a combination 
of the variation of the intrinsic colors of the background stars, and the 
departures from spherical symmetry of this model. At large radii, the former 
effect dominates, while at small and medium radii, the latter effect is more 
important.}
\end{figure}

\clearpage
\begin{figure}
\figurenum{7}
\epsscale{0.70}
\plotone{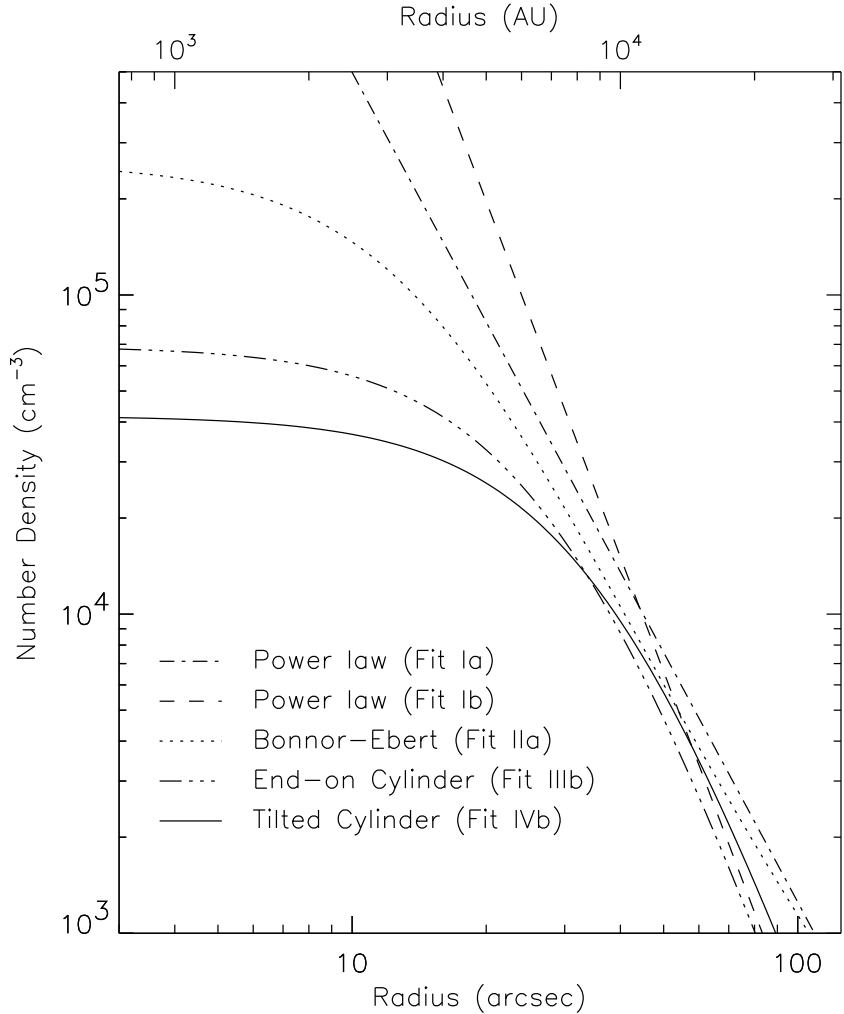}
\caption{Number density of molecular hydrogen vs.\ radius for the various
best fit models of L694--2 from Table~1, assuming mean molecular weight of
$\mu=2.29$, and a hydrogen mass fraction $X_{{\rm H}}=0.73$. 
The two power-law models 
({\em dash-dot} and {\em dashed}) have been normalized to the same column 
density as the Bonnor-Ebert model ({\em dotted}). For the
Bonnor-Ebert model and the two Cylindrical models ({\em dash-dot-dot-dot} 
and {\em solid}), the normalization is predicted directly
from the fitted profile shape. Assuming the standard gas-to-dust ratio,
reddening law, distance, etc.\, the Bonnor-Ebert model must be scaled by a
factor of $\cal{F}={}$4.3 to match the observed color excess. In the context
cylindrical models this scaling can be interpreted in terms of the 
extension of the cylinder along the line of sight. The extinction 
measurements cannot penetrate the region $r \leq 30''$ (7500~AU) where the 
various density profiles diverge from each other. The structure in this 
region can be probed with interferometer observations of dust emission.}
\end{figure}

\end{document}